\documentclass[a4paper,11pt]{article}
\pdfoutput=1 

\usepackage{jcappub} 

\usepackage{enumitem}

\title{North-South non-Gaussian asymmetry in Planck CMB maps}

\author[a]{A. Bernui}
\author[b]{A. F. Oliveira}
\author[c]{T. S. Pereira}

\affiliation[a]{Observat\'orio Nacional, Rua General Jos\'e Cristino 77, 
S\~ao Crist\'ov\~ao, 20921-400 Rio de Janeiro -- RJ, Brazil}
\affiliation[b]{Instituto de F\'{\i}sica e Qu\'{\i}mica, 
Universidade Federal de Itajub\'a, 37500-903 Itajub\'a -- MG, Brazil}
\affiliation[c]{Departamento de F\'{\i}sica, Universidade Estadual de Londrina, 
Rod. Celso Garcia Cid, Km 380, 86057-970, Londrina -- PR, Brazil}

\emailAdd{abernui@gmail.com}
\emailAdd{adhimar@unifei.edu.br}
\emailAdd{tspereira@uel.br}

\abstract{ 
We report the results of a statistical analysis performed with the four foreground-cleaned Planck maps by means of a suitably defined {\it local-variance} estimator. Our analysis shows a clear dipolar structure in Planck's {\it variance map} pointing in the direction $(l,b) \,\simeq\, (220^{\circ},-32^{\circ})$, thus consistent with the North-South asymmetry phenomenon. Surprisingly, and contrary to previous findings, removing the CMB quadrupole and octopole makes the asymmetry stronger. 
Our results show a maximal statistical significance, of $98.1\%$ CL, in the scales ranging from $\ell=4$ to $\ell=500$. 
Additionally, through exhaustive analyses of the four foreground-cleaned and individual frequency Planck maps, we find unlikely that residual foregrounds could be causing this dipole variance asymmetry. 
Moreover, we find that the dipole gets lower amplitudes for larger masks, evidencing that most of the contribution to the variance dipole comes from a region near the galactic plane. Finally, our results are robust against different foreground cleaning procedures, different Planck masks, pixelization parameters, and the addition of inhomogeneous real noise.}

\begin{document}
\maketitle
\flushbottom

\section{Introduction} \label{introduction}

The temperature fluctuations of the Cosmic Microwave Background radiation (CMB), 
recently released by the {\em Planck collaboration}~\cite{PLA-I}, confirmed with outstanding precision the concordance cosmological model, $\Lambda$CDM~\cite{PLA-XV,PLA-XVI,WMAP9}. 
Such exquisite set of cosmological information allows us to test two fundamental properties 
of the universe expected after the standard inflationary 
phase~\cite{Bartolo04,Komatsu1,Bassett2006,Linde2008}, namely that the CMB field is, at 
large-angles, nearly Gaussian and statistically isotropic 
(see, e.g.,~\cite{Abramo10,PLA-XXIII} and refs. therein). 

Previous studies using WMAP data indicate significant departure from either gaussianity or 
statistical isotropy at the largest angular scales -- an unexpected result in the $\Lambda$CDM 
model~\cite{Abramo06,Abramo09,Aghanim:2013suk,Pereira09,Bernui06,Bernui07,Bernui2008b,Bernui2009a,Bielewicz04,Copi04a,Copi04b,
Copi06,Copi07,Copi10,Copi13a,Copi13b,Cruz05,
Eriksen04,Eriksen07,Gordon,Gruppuso10,Gruppuso11,Mandolesi,
Hansen04a,Hansen04b,Huterer,Jaffe,Kahniashvili,Koivisto,Land05,Land07,OCTZ,TOCH,Paci,Rath,%
Samal08,Samal09,Vielva04,Vielva06,Vielva07,Urban,Zhao,%
Fabre,Hansen12,Kashino,Rath13,Rassat14}, though possibly disputable \cite{WMAP7}. 
These phenomena, also called {\it anomalies}, 
have been now confirmed with similar high confidence levels, $\sim 3 \sigma$, by the 
{\em Planck collaboration} with CMB foreground-cleaned maps~\cite{PLA-XXIII}. 
On the other side, only small magnitude Gaussian deviations from primordial origin have been 
detected in Planck data~\cite{PLA-XXII,PLA-XXIV}. 
However, there are more potential sources of non-Gaussianity (NG) in the CMB data than 
just primordial NG~\cite{Komatsu2,PNG-Liguori,Chen-2010,WMAP-7yr-Jarosik,%
Su-Yadav,Komatsu03}. 
These include galactic foregrounds remnants and secondary anisotropies coming from 
processes after the last scattering 
surface~\cite{WMAP-7yr-Gold,PLA-XXIV,Munshi,Aghanim,Novaes12,%
Chiang03,Naselsky,Delabrouille08,Novaes,Saha,Pietrobon09,Pietrobon10a,%
Pietrobon10b,Pratten,Smith,Vielva09,Zhao12}. 
In particular, Gaussian analyses for large angular scales are delicate because galactic 
foregrounds contaminations are not completely understood and, as a consequence, galactic 
cut-sky masks are still necessary in CMB data analyses~\cite{PLA-XXIV}. 
Monteser\'{\i}n et al. (2008)~\cite{Monteserin08} reported an anomalously low variance 
distribution in WMAP3 maps at $98.7\%$~CL. 
Cruz et al. (2011)\cite{Cruz11} confirmed this result in WMAP5 and WMAP7 data, also pointing that some regions near the galactic plane present an anomalously high variance ($95.6\%$ CL) in the south ecliptic hemisphere. Their analyses, using various galactic cut-sky masks, suggest that foreground residuals could explain the results, besides a possible connection with the CMB quadrupole-octopole alignment was 
investigated. Gruppuso et al. (2013) \cite{Gruppuso13}, using a different estimator, also found a low variance at large scales in WMAP9 data, basically in agreement with~\cite{Monteserin08,Cruz11}. 
%
%
More recently, the {\em Planck collaboration}~\cite{PLA-XXIII} and Akrami et al. 
(2014)~\cite{Akrami} studied the local variance in hemispheres and disks finding again an anomalous high variance in the south ecliptic hemisphere. 

In recent works~\cite{BR09,BR10,BR12} one of us have proposed two large-angle NG 
estimators based on skewness and kurtosis momenta performed on spherical caps on the 
CMB sphere. We found that this directional mapping approach is suitable when a cut-sky mask has to be used because it minimizes the effect of incomplete data in the CMB sky. 
These indicators provide a directional map of local NG due to its possible non-uniform distribution in the CMB maps, also giving information about the angular scale dependence of such contributions. Results obtained in previous analyses~\cite{BR09,BR10} using WMAP maps suggest that the NG captured there is not of primordial origin, although it might have a primordial component.

The aim of the present work is to conduct an analysis of the local variance in Planck 
foreground-cleaned maps, using a prescription similar to that of Refs.~\cite{BR09,BR10,BR12}. For this we implement a simple estimator of statistical variance, applying it to patches of the CMB sky. 
The information from all the patches is then used to produce an associated {\em Variance}-map, or simply $V\!-$map, which contains the signatures of the analysed CMB map. Our analyses investigate the possibility that foreground remnants in the galactic region could be the source of departures from Gaussianity and statistical  
isotropy, by considering several cut-sky Planck masks and three frequency band Planck 
maps, in addition to the four foreground-cleaned maps. To calculate the confidence level of our results we shall compare properties of these $V\!-$maps from Planck data with $V\!-$maps from simulated Monte Carlo (MC) CMB maps. These maps are obtained as Gaussian and statistically isotropic realisations from a seed angular power spectrum corresponding to the $\Lambda$CDM concordance model. Accordingly, the masking procedure applied to Planck CMB data is also applied to the MC maps. 

In section~\ref{pla-maps} we briefly review the main features of the four foreground-cleaned {\em Planck} maps and the masks to be used in the analyses. 
In section~\ref{method} we describe our variance estimator and explain the methodology to study the statistical Gaussian and isotropy attributes of Planck maps. 
The procedure delineated in this section will be used, in section~\ref{results}, to investigate directional large-angle deviations from the standard statistical scenario of the Planck data as compared with simulated maps. Our analysis includes realistic features of the Planck data, like their inhomogeneous noise maps and galactic cut-sky masks. Finally, in section~\ref{conclusion}, we summarize our main results, present our conclusions and 
final remarks.

\section{Foreground-reduced Planck maps} \label{pla-maps}

The Planck satellite observed the sky in nine frequency bands, from 30 to 857 
GHz~\cite{PLA-I,PLA-XII}. The use of four {\em component separation techniques}, which efficiently identifies the sources of contaminating emissions present in the data set, have allowed the {\em Planck collaboration} to produce four high resolution and almost full sky foreground-cleaned CMB maps~\cite{PLA-XII}. They are: the Spectral Matching Independent Component Analysis ({\sc smica})~\cite{Cardoso08}, the Needlet Internal Linear Combination ({\sc nilc})~\cite{Delabrouille08}, the Internal Template 
Fitting Spectral Estimation Via Expectation Maximization ({\sc sevem})~\cite{Fernandez-Cobos12}, and the combined approach termed Commander-Ruler ({\sc CR})~\cite{Eriksen06,PLA-XII}. 

Each of the foreground cleaning methods provides a CMB map with its Component Separation Confidence mask --also termed {\it validation} mask or simply {\sc val}mask-- outside which the corresponding CMB data is considered to be foreground-cleaned, and also a noise map containing an estimate of the real inhomogeneous pixel's noise. In addition, the {\sc smica} and {\sc nilc} maps were released with their own {\it inpainting} mask, or simply {\sc inp}mask. Regarding the masks, there also exist the separation component minimum mask, termed M82, and the U73 mask, which is the union of the confidence galactic masks plus the point sources 
mask~\cite{PLA-XII} (see Table~\ref{table1}).  

The effect of realistic anisotropic noise, due to the different number of times that each pixel was observed by the probe, is taken into account according to the specifications of each foreground-cleaned method. 
For this, each component separation procedure provides an estimate of the pixel's noise in the output CMB map, information released together with each foreground-cleaned Planck map as a full-sky map, termed {\em noise map}~\cite{PLA-XII,PLA-XXIII,PLA-XXIV}. Thus, we use the noise maps in a ``signal + noise'' analysis to find their effect on $V\!-$maps, i.e., we apply our estimator after adding a noise map to its corresponding foreground-cleaned Planck map. In section 4 we consider the analyses with and without including this realistic anisotropic noise component, which is done by adding to the foreground-cleaned map its corresponding noise map, and also considering different Planck masks. 

\begin{table}[h] 
\begin{center}
\begin{tabular}{lc} 
\hline \hline
\ \ \ \  Planck mask \ \ & \ \ \ \ \ \ \ \ \ \ \ \ \ \ $f_{\mbox{sky}}$ \ \ \ \ \ \\ 
\hline
\ \ \ {\sc smica}~--~{\sc inpmask}   & \ \ \ \ \ \ \ \ \ 0.97   \\
\ \ \ {\sc smica}~--~{\sc valmask}   & \ \ \ \ \ \ \ \ \ 0.89   \\
\ \ \ {\sc nilc}~--~{\sc inpmask}       & \ \ \ \ \ \ \ \ \ 0.97   \\
\ \ \ {\sc nilc}~--~{\sc valmask}       & \ \ \ \ \ \ \ \ \ 0.93   \\
\ \ \ {\sc sevem}~--~{\sc valmask}  & \ \ \ \ \ \ \ \ \ 0.76   \\
\ \ \ {\sc CR}~--~{\sc valmask}          & \ \ \ \ \ \ \ \ \ 0.75   \\
\ \ \ M82~--~ {\sc minimal mask}    & \ \ \ \ \ \ \ \ \ 0.82   \\
\ \ \ U73~--~ {\sc union mask}        & \ \ \ \ \ \ \ \ \ 0.73   \\
\hline \hline
\end{tabular}
\end{center}
\caption {The $f_{\mbox{\footnotesize sky}}$ values give the available sky fraction of 
a CMB map when a Planck mask is applied to it.} 
\label{table1}
\end{table}

\section{The variance estimator} \label{method} 

We start this section by explaining the procedure for constructing the variance map ($V\!-$map) of a 
given CMB map. Let $\Omega_j \equiv \Omega(\theta_j,\phi_j) \in \mathbb{S}^2$ be a hemisphere on 
the celestial sphere, with center at the $j^{\,\rm{th}}$ pixel, $j=1, \ldots, N_{\mbox{\footnotesize hem}}$, 
where $(\theta_j,\phi_j)$ are the angular coordinates of the $j^{\,\rm{th}}$ pixel, and 
$N_{\mbox{\footnotesize hem}}$ is the number of hemispheres. 
The number of hemispheres and the coordinates of their centers are defined using the HEALPix 
pixelization scheme~\cite{Gorski05}. 
Moreover, the hemisphere's centers are uniformly distributed on $\mathbb{S}^2$ 
and the union of them completely covers the celestial sphere. 

The variance of the data inside each hemisphere can be calculated simply by
\begin{eqnarray} \label{V-map}
&&V_j   =  \frac{1}{ n_{\rm p} } \sum_{i=1}^{n_{\rm p}}
\left(\, T_j^i\, - \overline{T_j} \,\right)^2 \, , \label{Vdef} 
\end{eqnarray} 
where $n_{\rm p}$ is the number of pixels in the $j^{\,\rm{th}}$ hemisphere, $T_j^i$ is the temperature fluctuation at the $i^{\,\rm{th}}$ pixel and $\overline{T_j}$ is the mean CMB temperature fluctuation of the $j^{\,\rm{th}}$ hemisphere.

The values $V_j$ obtained in this way give a local measure of the variance in the direction 
$(\theta_j, \phi_j)$. Patching together the set of values $\{ V_j, \, j=1,...,N_{\mbox{\footnotesize hem}} \}$ in a sphere with $N_{\mbox{\footnotesize hem}}$ pixels we obtain a colored (pixelized) celestial sphere. The Mollweide projection of this sphere is termed the $V\!-$map: it is the final product of the application of our variance estimator to a given CMB map. According to the scale of colors, the minimum (maximum) value of the set $\{ V_j \}$ corresponds to the bluest (reddest) pixel. 
With the above prescription one can obtain a quantitative measure of anomaly of a real map by simply calculating the angular power spectrum of its corresponding $V\!-$map, and then comparing it with the mean power spectra obtained from MC simulations.

Because the $V\!-$map assigns a real value to each pixel in the celestial sphere 
$\mathbb{S}^2$, that is $V = V(\theta,\phi)$, one can expand it in spherical harmonics: 
$V(\theta,\phi) = \sum_{L, M} A_{L M} Y_{L M}(\theta,\phi)$, 
where the set of values $\{ v_{L},\, L = 0,1,2,\cdots \}$, given by 
\begin{eqnarray} \label{V-spectrum}
v_{L} \,\equiv\, \frac{1}{2 L + 1}\, \sum_{M={\mbox{\small -}}L}^{L} \, |A_{L M}|^2 \, , 
\end{eqnarray}
is the angular power spectrum of the $V\!-$map. 
Given that we are interested in the large-scale NG deviations, we shall concentrate on the low-$L$ angular power spectrum $\{ v_{L}, \,\mbox{for}\,\, L = 1,2,\cdots,10 \}$ of the $V\!-$map. 

Before proceeding, let us clarify some points of our whole prescription. In what concerns the implementation of a local variance estimator, note that Eq. (\ref{V-map}) is the simplest mathematical possibility. On the other hand, the choice of spherical caps with an aperture of $90^{\circ}$ is motivated by the fact that, when scanning a map to calculate its $V\!-$map, the procedure considers caps whose centers are close or even within the masked region. In these cases, the variance of the data inside the cap is performed with a smaller number of pixels, which introduces additional statistical noise as compared to caps away from the masked region. As it turns out,
this effect can be minimized by choosing spherical caps having aperture of $90^{\circ}$, that is, hemispheres~\cite{BR09,BR10,BR12}. Note that our prescription is thus different from the one adopted in~\cite{Akrami}, where not only the size of the caps is allowed to vary, but also masked pixels are excluded. As we will see, our results are compatible with their findings.

In the next section we use the above procedure to generate $V\!-$maps from the set of 1,000 Gaussian and statistically isotropic simulated maps -- from now on called 
$V^{\mbox{\footnotesize\sc g}}\!-$maps -- from which we obtain the corresponding spectra, $v_L^{\mbox{\footnotesize\sc g}}$, and mean values, $\overline{v}_{L}^{\mbox{\footnotesize\sc g}}$. We then compare the spectra of the $V\!-$maps produced from the Planck maps, from now on called $V^{\mbox{\footnotesize\sc pla}}\!-$maps, with the mean value $\overline{v}_{L}^{\mbox{\footnotesize\sc g}}$. We finally emphasize that, although it might be clear from the context, the $V^{\mbox{\footnotesize\sc g}}\!-$maps themselves are not normally distributed.

\subsection{Statistically Isotropic Gaussian maps} \label{MC-maps}

\begin{figure}
\includegraphics[angle=90,scale=0.3]
{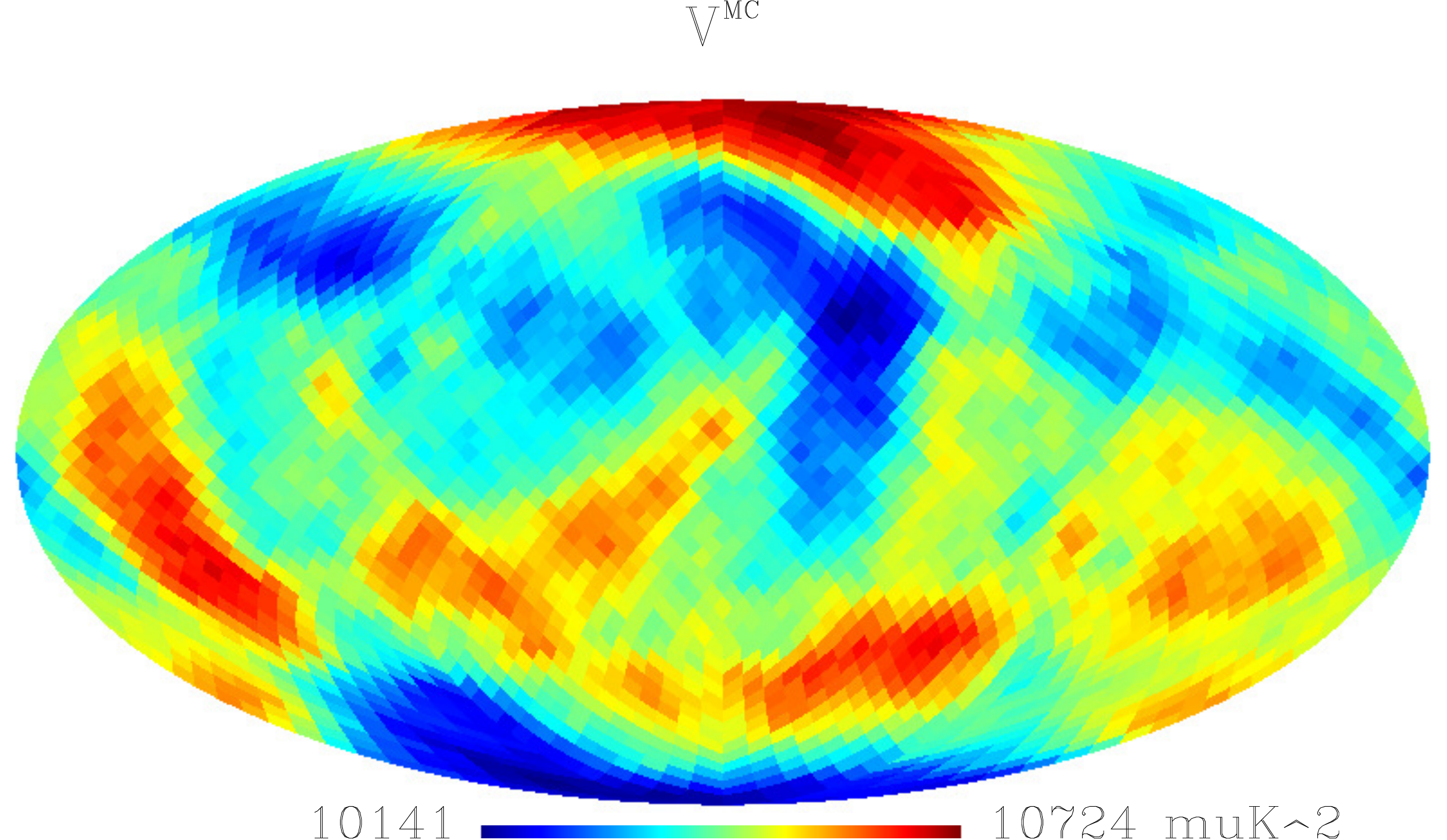}
\includegraphics[angle=90,scale=0.3]
{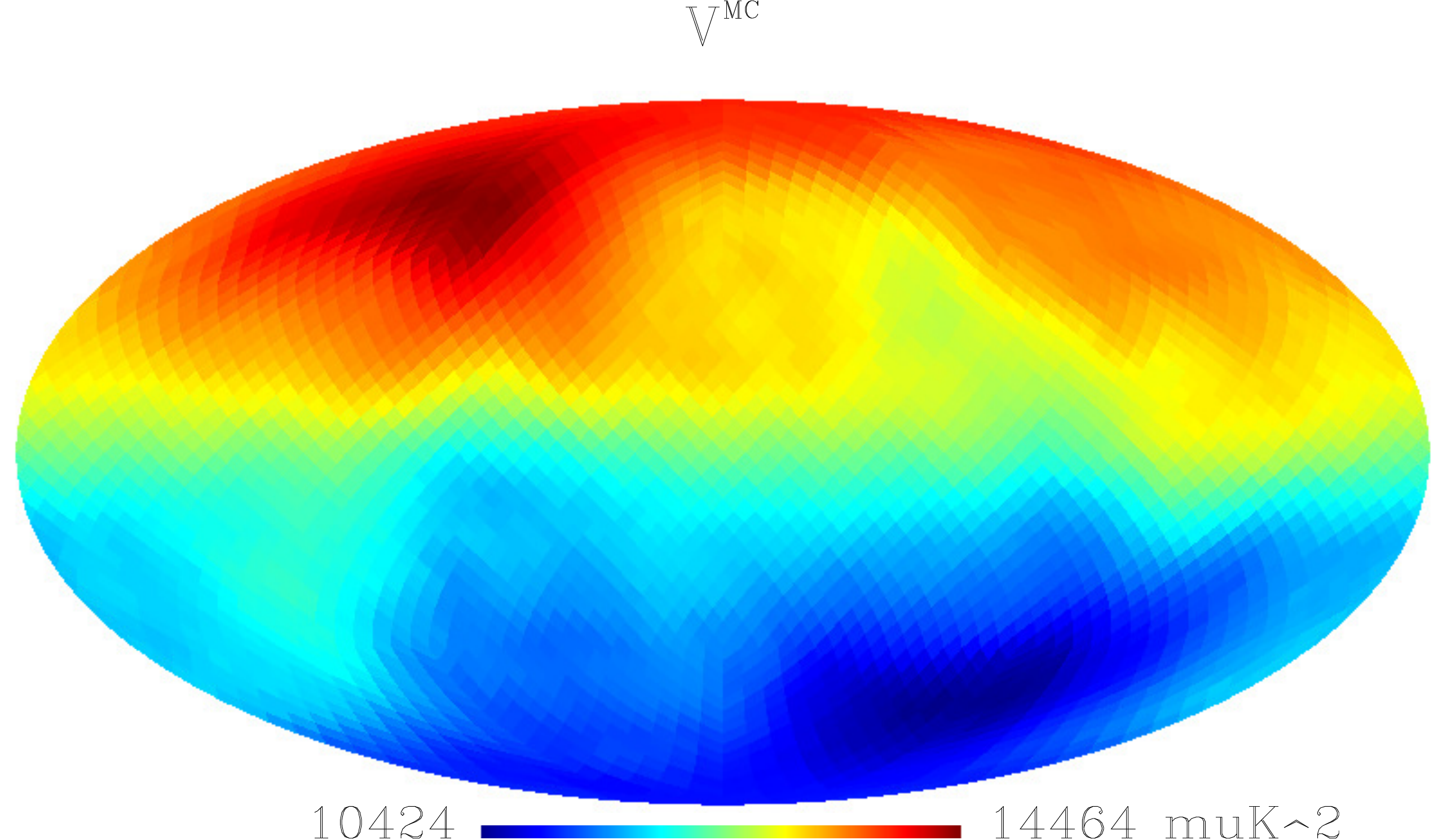}
\caption{Variance maps from Gaussian and statistically isotropic simulations. 
We show $V^{\mbox{\footnotesize\sc g}}\!-$maps with the minimum (left, smaller than the data) and the 
maximum (right, larger than the data) variance dipole values produced from the set of 1,000 MC CMB maps, 
in units of $\mu{\rm K}^2$.} 
\label{fig1}
\end{figure}

Our Gaussian and statistically isotropic MC maps were obtained as random realizations 
from a seed angular power spectrum, which corresponds to the $\Lambda$CDM concordance 
model~\cite{PLA-I}, and the map-making process is performed using the {\sc synfast} facility from the HEALPix package~\cite{Gorski05}. We test the robustness of our outcomes with two different angular resolutions of the 
$V\!-$maps, that is, with $N_{\mbox{\footnotesize hem}} = 768$ and with 
$N_{\mbox{\footnotesize hem}} = 3,\!072$. 
For illustration, we show in Fig.~\ref{fig1} two representative 
$V^{\mbox{\footnotesize\sc g}}\!-$maps from a MC simulations; notice in these figures the minimum and the maximum values. 

In order to generate a $V\!-$map, either from a MC or foreground-cleaned Planck map, we first choose the angular scales to be included in analysis, $\ell \in [\ell_\text{min},\ell_\text{max}]$, which in turn determine the angular resolution ($N_{\mbox{\footnotesize side}}$) of the map. Planck and simulated CMB maps are analysed under the same conditions: mask, angular-scale's interval, $N_{\mbox{\footnotesize side}}$, and $N_{\mbox{\footnotesize hem}}$. 
Thus, given the set of 1,000 MC maps, we produce their corresponding 1,000 
$V^{\mbox{\footnotesize\sc g}}\!-$maps and calculate their associated power spectra, namely, 
$\{ \{ v_{\,L} \}^{\,\mathbf{i} } \}$, for $\mathbf{i} = 1, \cdots, 1,\!000$ and $\,L=1, \cdots, 10\,$. 
Finally, we compute the mean angular power spectra of the 
$V^{\mbox{\footnotesize\sc g}}\!-$maps 
\begin{eqnarray} \label{mean-spectra}
\overline{v}_{L}^{\mbox{\footnotesize\sc g}} \,=\, 
\frac{1}{1000} \,\sum_{\mathbf{i} = 1}^{1000} v_{\,L}^{\,\mathbf{i}} \, . 
\end{eqnarray}
These values are then used to obtain the statistical significance 
(i.e., the goodness-of-fit) of the spectrum $v_{L}^{\mbox{\footnotesize {\sc pla}}}$, obtained from the $V^{\mbox{\footnotesize\sc pla}}\!-$map.

\section{Statistical estimator applied to Planck maps} \label{results}

In this section we perform variance analyses of the four foreground-cleaned Planck maps. 
We first calculate their angular power spectra $v_{L}^{\mbox{\footnotesize\sc pla}}$, and the 
corresponding statistical confidence level by comparison with the mean spectra 
$\overline{v}_{L}^{\mbox{\footnotesize\sc g}}$. 
We find that the dipolar term of the $v_{L}^{\mbox{\footnotesize\sc pla}}$'s spectra is the dominant term, and appears to be robust under foreground-cleaning procedures (i.e., a similar result is obtained for the four Planck maps), cut-sky masks, inhomogeneous pixel's noise, and different estimator's parameter $N_{\mbox{\footnotesize hem}}$. Then we investigate several angular scale's intervals looking for the origin of this variance dipole phenomenon. At the end of this section we discuss the possibility that residual foregrounds could be causing it.

\subsection{Angular power spectra analyses of $V$-maps}%
\label{planck-maps}

The CMB maps analysed in this subsection contain the angular-scales 
$\ell \in[2,\, 1,\!000]$. We shall study two cases: the `pure signal' case and the `signal + inhomogeneous noise' case. Initially we use $N_{\mbox{\footnotesize side}} = 512$, 
$N_{\mbox{\footnotesize hem}} = 3,\!072$, but to validate our results of this subsection we also consider other pixelization scheme's parameters as robustness tests. 

\begin{figure*} 
\begin{center}
\includegraphics[scale=0.29,angle=90]{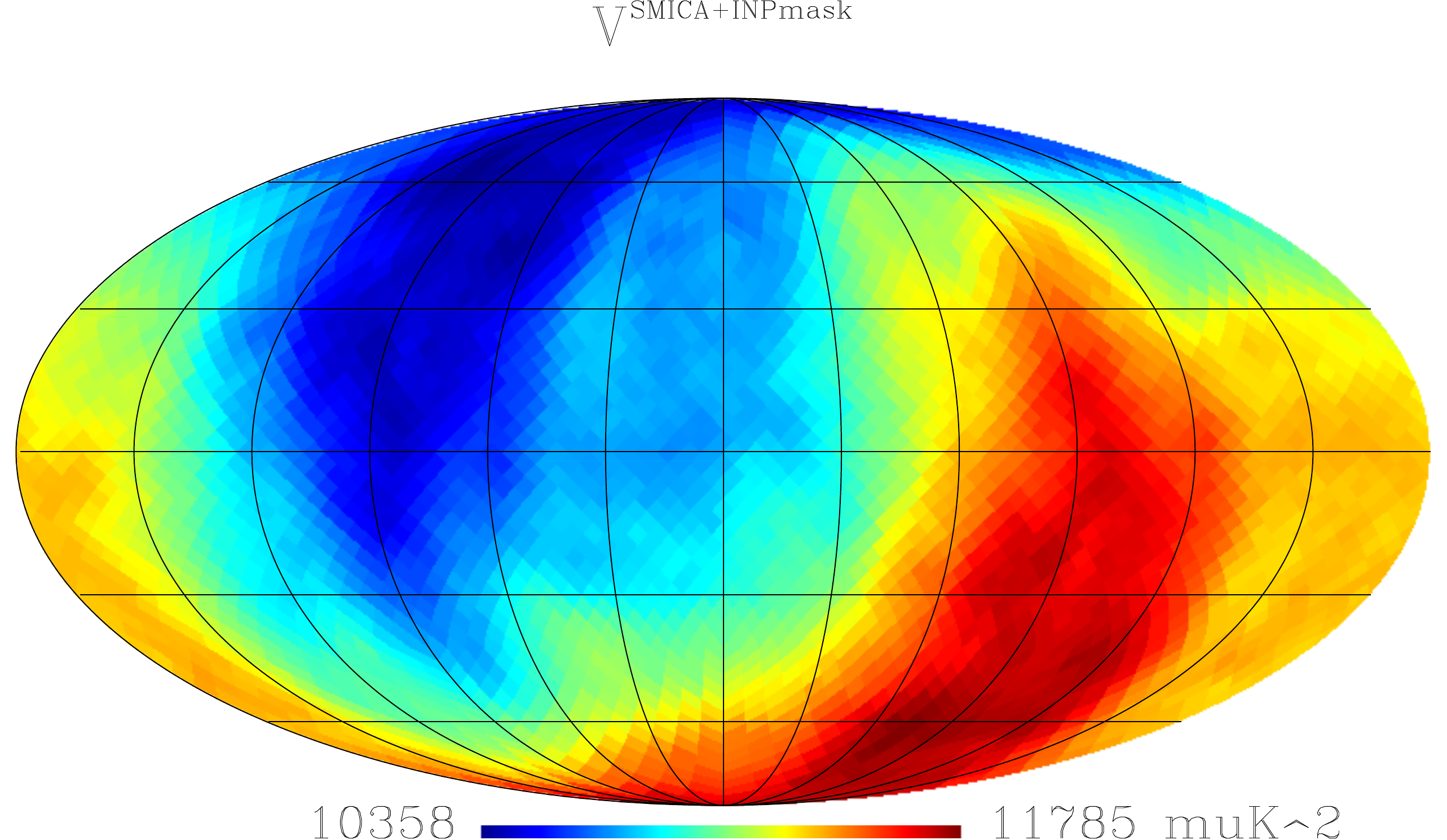}\vspace{2mm}
\includegraphics[scale=0.29,angle=90]{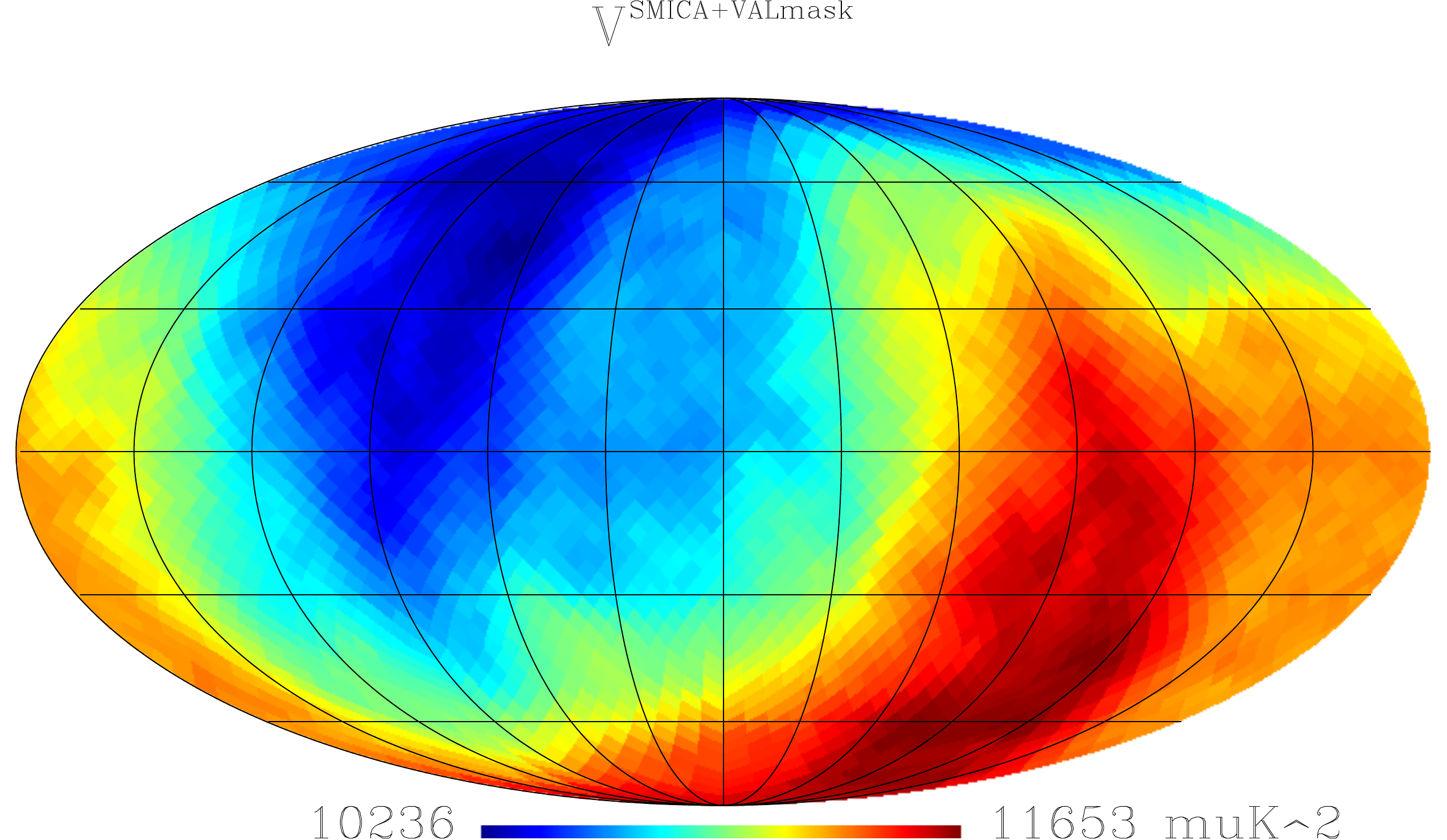}
\includegraphics[scale=0.29,angle=90]{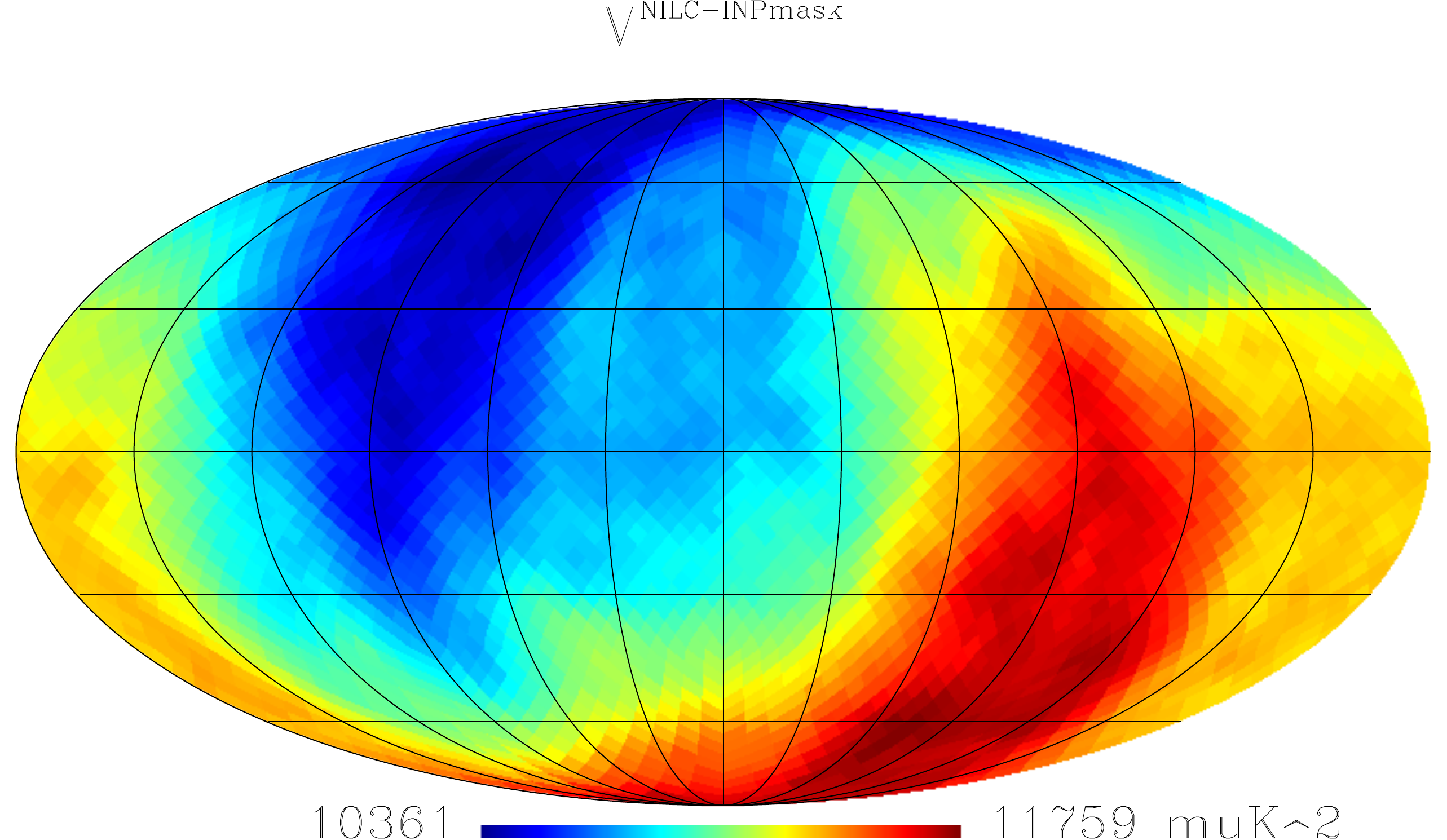}\vspace{2mm}
\includegraphics[scale=0.29,angle=90]{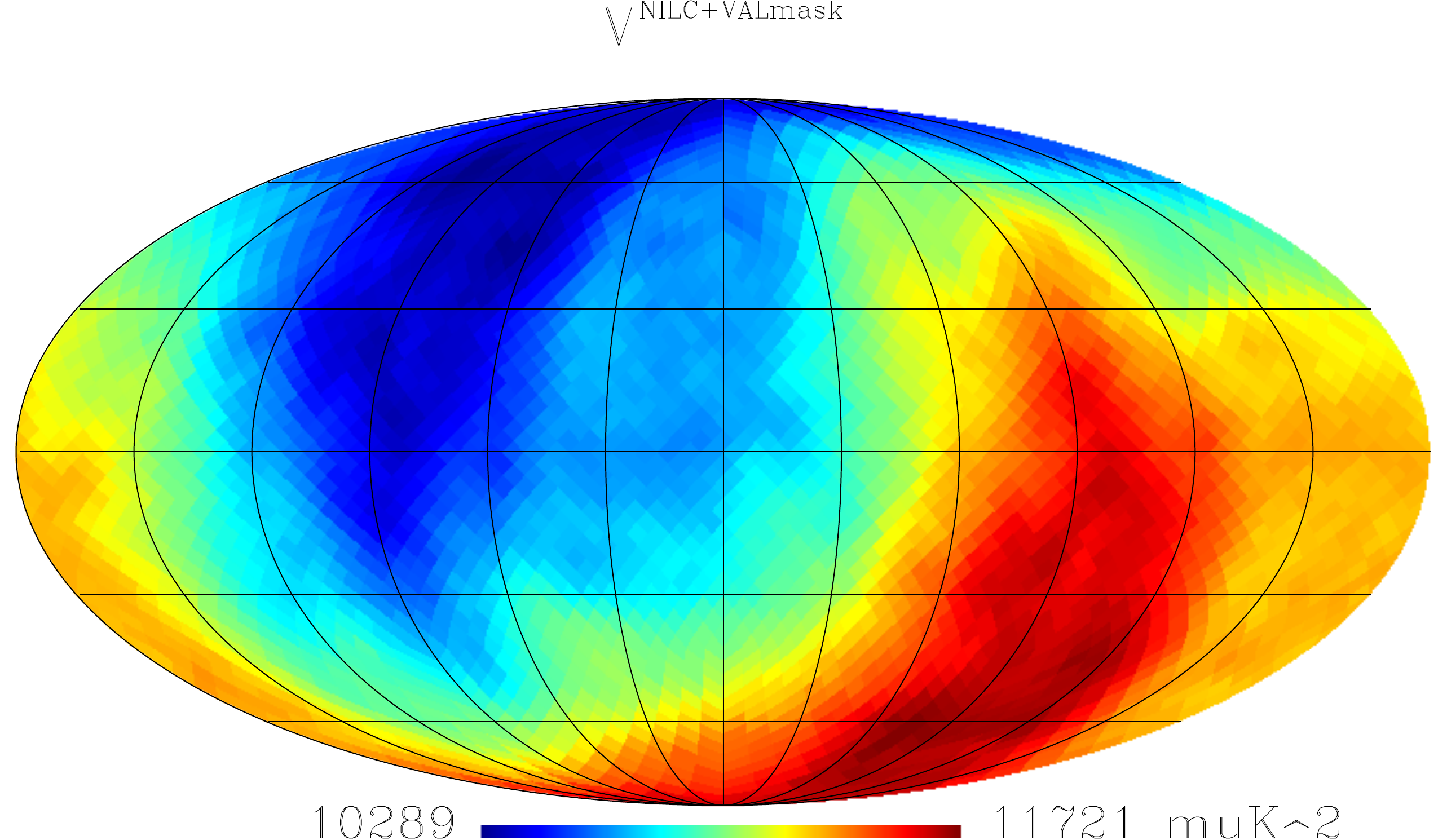}
\includegraphics[scale=0.29,angle=90]{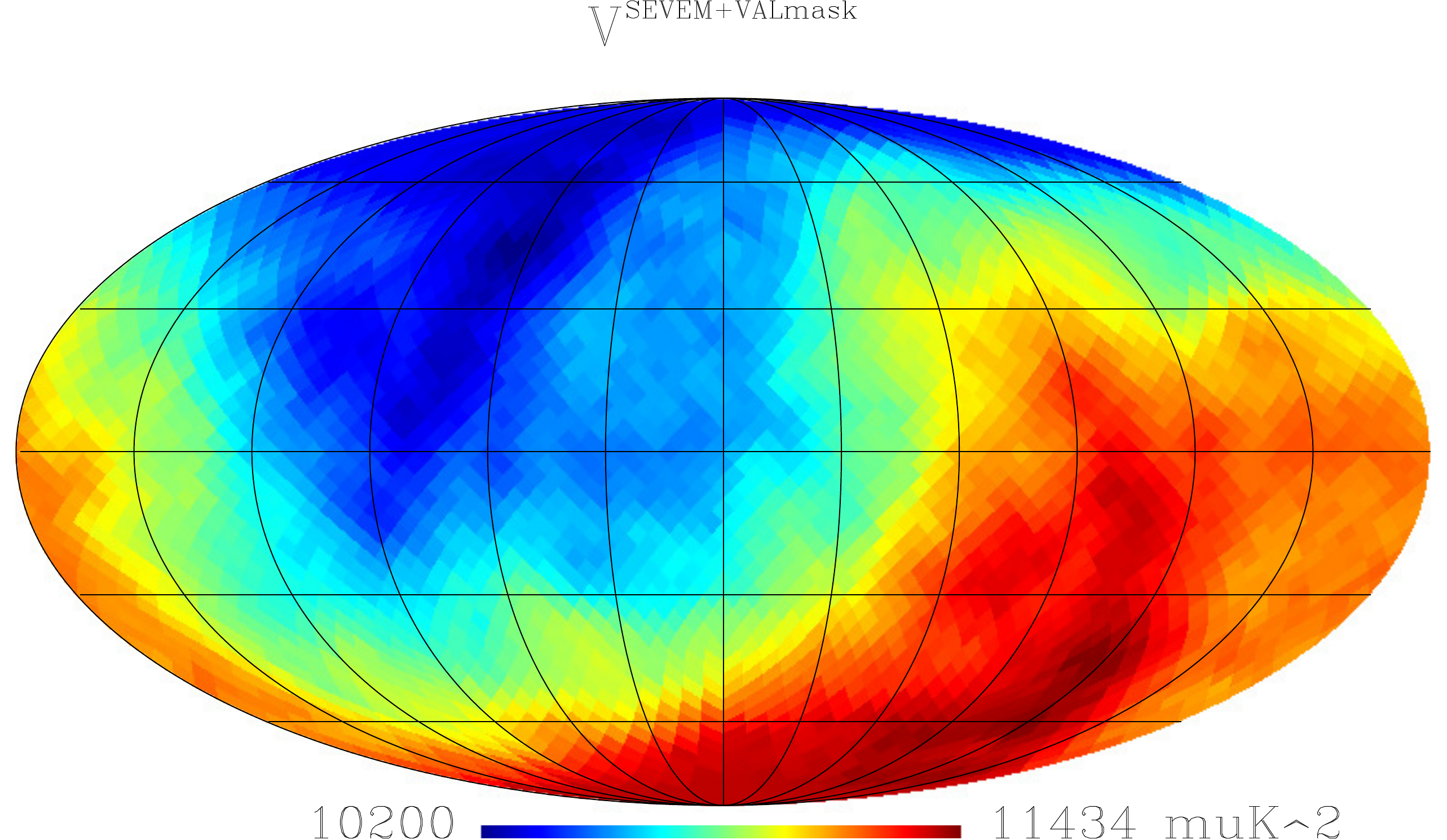}
\includegraphics[scale=0.29,angle=90]{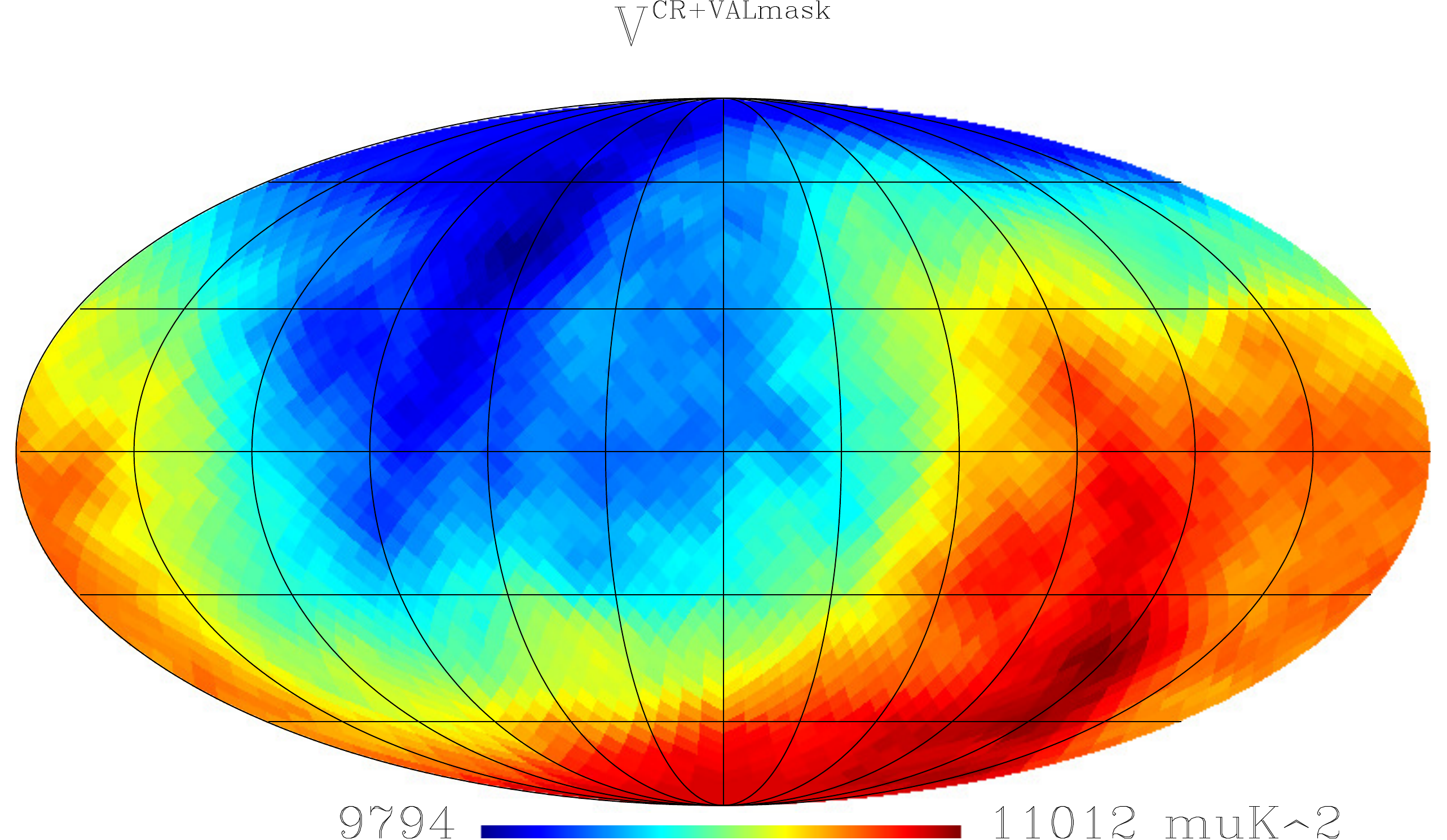}
\caption{\label{fig2} 
Variance maps of four different foreground-cleaned Planck maps. 
In the top row we have the $V\!-$maps obtained from the foreground-cleaned 
{\sc smica} Planck map using their {\sc inp}mask (left) and {\sc val}mask 
(right). Similarly, in the middle row we show the $V\!-$maps obtained from the foreground-cleaned {\sc nilc} Planck map using their {\sc inp}mask (left) and {\sc val}mask (right), respectively. Finally, in the bottom row we have the $V\!-$maps obtained from the foreground-cleaned {\sc sevem} (left) and {\sc CR} (right) Planck maps using their corresponding {\sc val}masks.}
\end{center}
\end{figure*}

The result of the application of our estimator on the four foreground-cleaned Planck maps can be observed in Fig.~\ref{fig2}, where we show the $V^{\mbox{\footnotesize\sc pla}}\!-$maps obtained using the four Planck maps and diverse masks. 
A common interesting feature noticed in these $V\!-$maps is the strong dipolar signal, independent of the maps and masks used to produce the $V^{\mbox{\footnotesize\sc pla}}\!-$map. Similar results are obtained in all the other cases investigated, which we do not show in Fig.~\ref{fig2} to avoid repetitions. 

\begin{figure}
\begin{center}
\mbox{\hspace{-1.cm}
\includegraphics[scale=0.52]%
{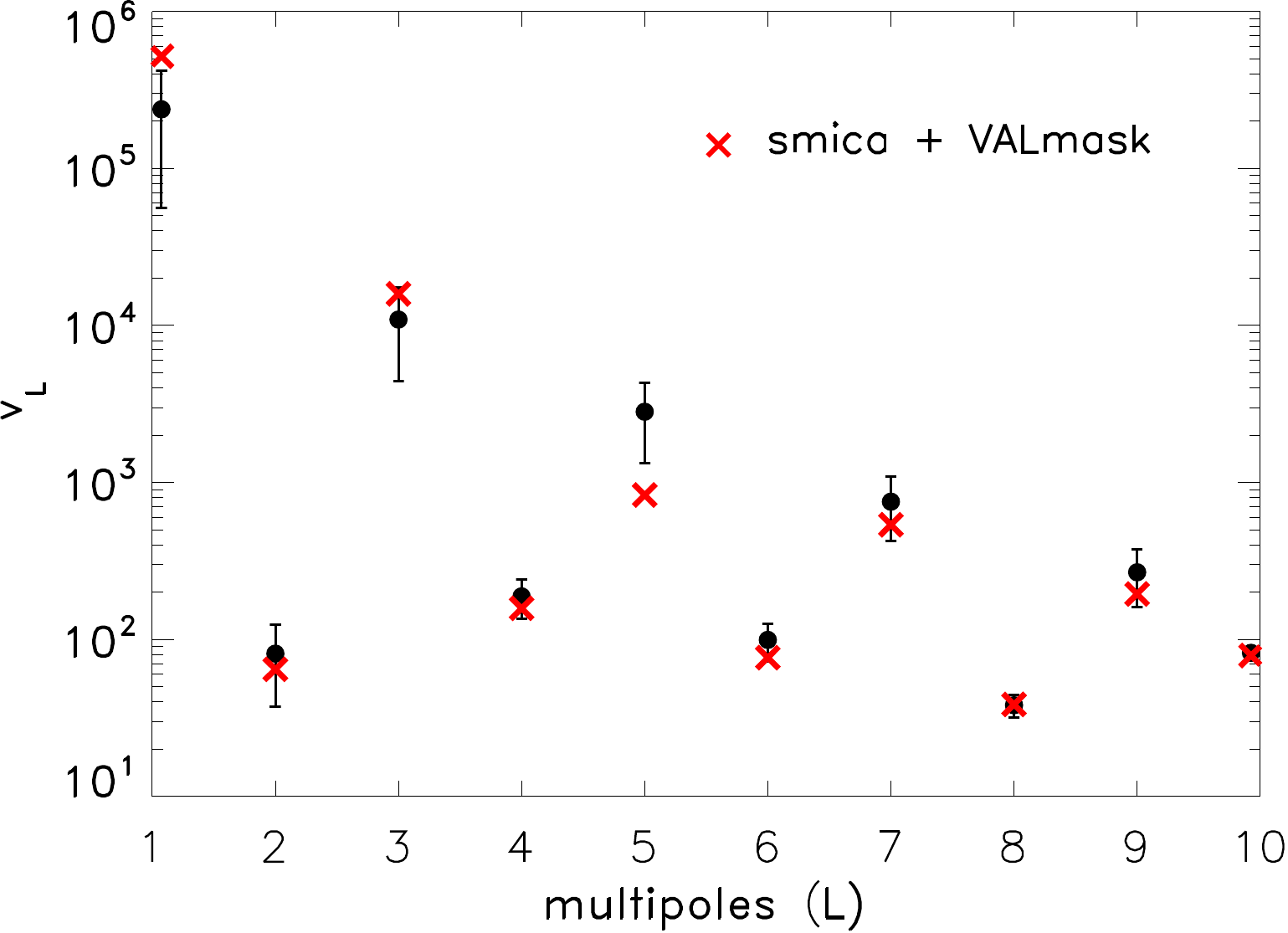}}
\caption{\label{fig3} 
Angular power spectrum of the $V^{\mbox{\footnotesize\sc smica}}\!-$map, generated 
applying our estimator to the {\sc smica} Planck map, after cutting the sky patch corresponding 
to its {\sc val}mask. 
The average spectrum $\overline{v}_{L}^{\mbox{\footnotesize\sc g}}$ generated from 
Gaussian and statistically isotropic CMB data is represented as dots with 1$\sigma$ 
error bars. The $\chi^2$ goodness-of-fit gives 7.3, for 9 d.o.f., corresponds to a $p-$value equal to 0.61, which lead us to conclude that this spectrum is fully consistent with the null hypothesis (i.e., a Gaussian and statistically isotropic universe). 
The $V^{\mbox{\footnotesize\sc smica}}\!-$map for this case, i.e., {\sc smica} map 
+ {\sc val}mask, is seen in the right panel of the first row in Fig.~\ref{fig2}, 
where the dipole variance points towards $(l, b) \,\simeq\, (245^{\circ},-35^{\circ})$. 
}
\end{center}
\end{figure}

In Fig.~\ref{fig3} we give a quantitative measure of the spectra of the 
$V^{\mbox{\footnotesize\sc pla}}\!-$maps as compared with the average spectra of the 
$V^{\mbox{\footnotesize\sc g}}\!-$maps (for simplicity we present only the {\sc smica} + {\sc val}mask case; the other cases show similar results). This comparison measures the possible departure of the Planck data with 
respect to the standard statistical scenario. In fact, an overall assessment of the statistical 
significance of the spectrum $v_{L}^{\mbox{\footnotesize {\sc pla}}}$ as compared with the average spectra 
$\overline{v}_{L}^{\mbox{\footnotesize\sc g}}$ data, given by the $\chi^2$ goodness-of-fit, supplies a 
measure of the statistical features present in Planck data. 
For instance, in Fig.~\ref{fig3} the goodness-of-fit test gives $\chi^2 \,=\,7.3$, for 9 d.o.f. (degrees of 
freedom), which means a good agreement between Planck data and Gaussian MCs, 
having in mind the large cosmic variance existent at these scales. 
Similar numerical analyses can be obtained for other cases.

As suggested by the Planck team~\cite{PLA-XXIV}, realistic experimental features such as galactic masks and anisotropic noise distribution should be used in Planck data analyses to test the robustness of the results. Moreover, because these noise maps have a large quadrupolar signal roughly aligned with the ecliptic, it is pertinent to evaluate the effect of such real anisotropic noise by comparing the cases with and without noise. For this, we calculate the $\chi^2$ values, for 9 d.o.f., considering the four foreground-cleaned Planck maps in different situations: considering the addition or not of their corresponding anisotropic noise maps and after that we apply the M82 mask or the U73 masks. Our results are shown in Table~\ref{table2}, where we emphasize that the $\chi^2$ values were not divided by the number of d.o.f. 
The conclusion is that the Planck data, after being cut with either M82 or U73 masks, and independent of the addition of the inhomogeneous pixel's noise, are fully consistent with the Gaussian and statistically isotropy hypotheses of the standard cosmological model.

\begin{table}[h]
\begin{center}
\begin{tabular}{lcc} 
\hline \hline
\ \ \ Planck map + mask \ \ 
& \ \ \ \ \ \ \ \ \ \ \ \ $\chi^2$ \ \ & \ \ \ \ \ \ \ \ \ $\chi^2_{\text \, inh. noise}$ \ \ \ \\ 
\hline
\ \ \ \ \ \ {\sc smica} + M82   & \ \ \ \ \ \ \ \ \ \ \ 8.6 \ \ \ & \ \ \ \ \ \ \ \ \ \ \ \ 8.2 \ \ \ \ \   \\
\ \ \ \ \ \ {\sc smica} + U73   & \ \ \ \ \ \ \ \ \ \ \ 8.8 \ \ \ & \ \ \ \ \ \ \ \ \ \ \ \ 8.6 \ \ \ \ \     \\
\ \ \ \ \ \ {\sc nilc} + M82       & \ \ \ \ \ \ \ \ \ \ \ 8.4 \ \ \ & \ \ \ \ \ \ \ \ \ \ \ \ 8.1 \ \ \ \ \     \\
\ \ \ \ \ \ {\sc nilc} + U73        & \ \ \ \ \ \ \ \ \ \ \ 9.0 \ \ \ & \ \ \ \ \ \ \ \ \ \ \ \ 8.7 \ \ \ \ \    \\
\ \ \ \ \ \ {\sc sevem} + M82  & \ \ \ \ \ \ \ \ \ \ \ 8.4 \ \ \ & \ \ \ \ \ \ \ \ \ \ \ \ 8.1 \ \ \ \ \   \\
\ \ \ \ \ \ {\sc sevem} + U73  & \ \ \ \ \ \ \ \ \ \ \ 8.6 \ \ \ & \ \ \ \ \ \ \ \ \ \ \ \ 8.3 \ \ \ \ \    \\
\ \ \ \ \ \ {\sc CR} + M82       & \ \ \ \ \ \ \ \ \ \ \ 11.6 \ \ \ & \ \ \ \ \ \ \ \ \ \ \ 11.1 \ \ \ \ \   \\
\ \ \ \ \ \ {\sc CR} + U73        & \ \ \ \ \ \ \ \ \ \ \ 12.9 \ \ \ & \ \ \ \ \ \ \ \ \ \ \ 12.4 \ \ \ \ \   \\
\hline \hline
\end{tabular}
\end{center}
\caption{$\chi^2$, for 9 d.o.f., 
obtained when each $\overline{v}_{L}^{\mbox{\footnotesize\sc pla}}$ spectrum is fitting the 
$\overline{v}_{L}^{\mbox{\footnotesize\sc g}}$ spectrum, considering the M82 and U73 masks, 
in two situations: `pure signal' case and `signal + inhomogeneous noise' case. 
The first (second) column correspond to the case without (with) the addition of 
inhomogeneous pixel's noise to the Planck maps before the variance analysis. 
Our results show a good agreement between the large-angle spectra of 
$V^{\mbox{\footnotesize\sc pla}}\!-$maps as compared with 
$V^{\mbox{\footnotesize\sc g}}\!-$maps. 
We stress that the $V^{\mbox{\footnotesize\sc g}}\!-$maps are not normally distributed.}
\label{table2}
\end{table}

One should note that the overall good agreement -- as evinced by the $\chi^2$ test 
(table~\ref{table2}) -- 
between the spectrum $v_{L}^{\mbox{\footnotesize\sc pla}}$ and 
$\overline{v}_{L}^{\mbox{\footnotesize\sc g}}$ for the scale's range $\,L=1, \cdots, 10\,$, 
does not exclude the possibility that a particular scale $L$ could be anomalous with respect to the Gaussian and statistically isotropic scenario. 
Indeed, one observes in Fig.~\ref{fig3}, that the dipole $v_{1}^{\mbox{\footnotesize\sc smica}}$ is the largest multipole value of the spectra, being $\sim$ 50 times greater than the sum of the other multipoles. This dominant dipole term, $L = 1$, similarly observed in the other three Planck maps, reflects what is being observed in the $V^{\mbox{\footnotesize\sc pla}}\!-$maps displayed in 
Fig.~\ref{fig2}. 
Moreover, this dipolar variance asymmetry seems to be related to the anomalous variance distribution found in WMAP maps~\cite{Cruz11}, and more recently in Planck 
data~\cite{PLA-XXIII,Akrami}. 
For these reasons, this suspicious dipolar phenomenon deserves detailed angular scale's analyses, which shall be done in the next subsection.

To end this subsection, we notice that these results are robust under the four Planck's foreground-cleaning procedures, the set of Planck masks~\ref{table2}, inhomogeneous pixel's noise (released by Planck's team), and pixelization scheme parameters (that is, $N_{\mbox{\footnotesize side}} = 256; \, 512$, 
$N_{\mbox{\footnotesize hem}}=768; \, 3,\!072$).

\subsection{Angular-scale analyses of the Variance dipole}%
\label{angular-scales}

\begin{figure}
\begin{center}
\includegraphics[scale=0.3,angle=90]{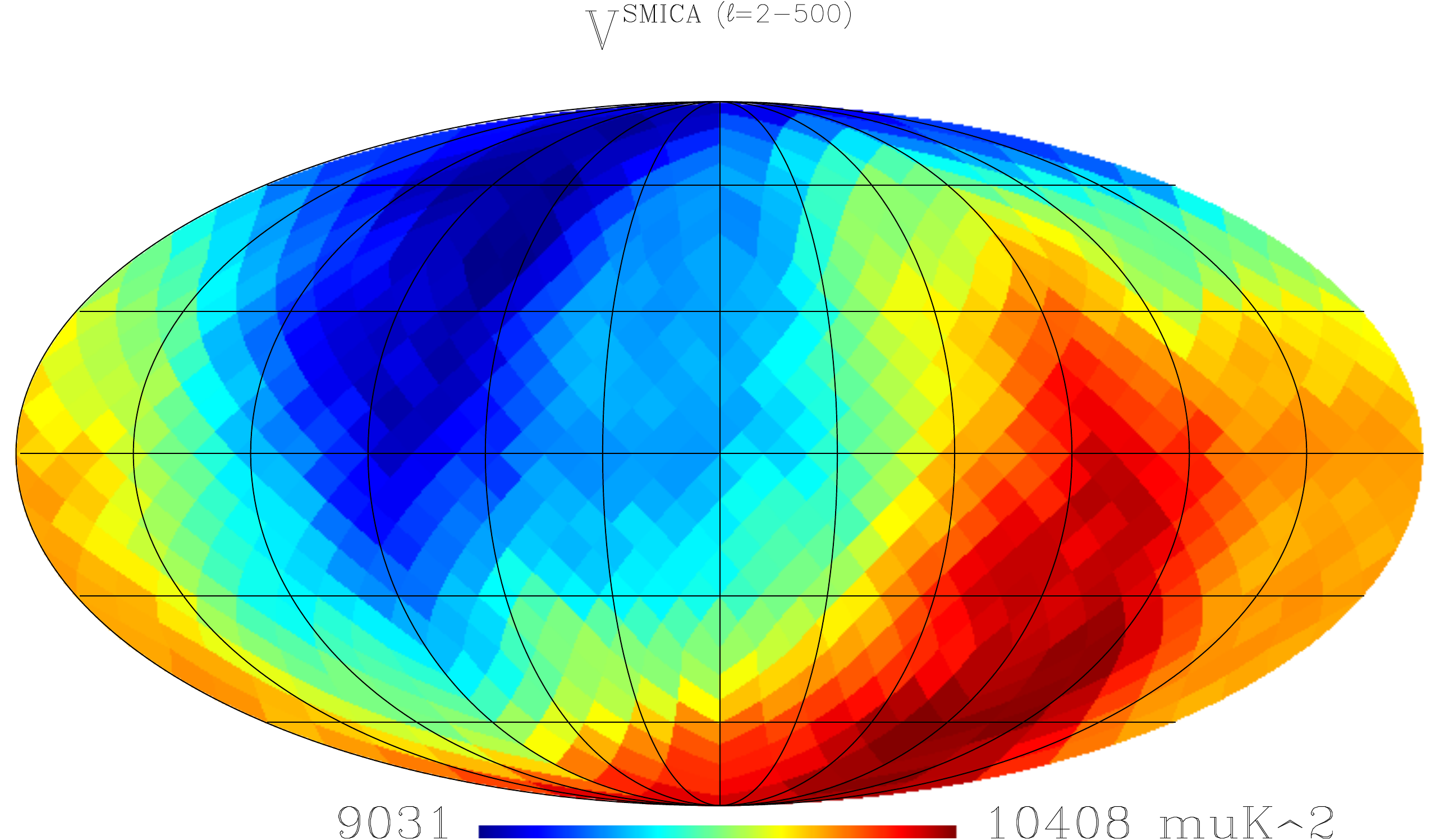}\vspace{1mm}
\includegraphics[scale=0.3,angle=90]{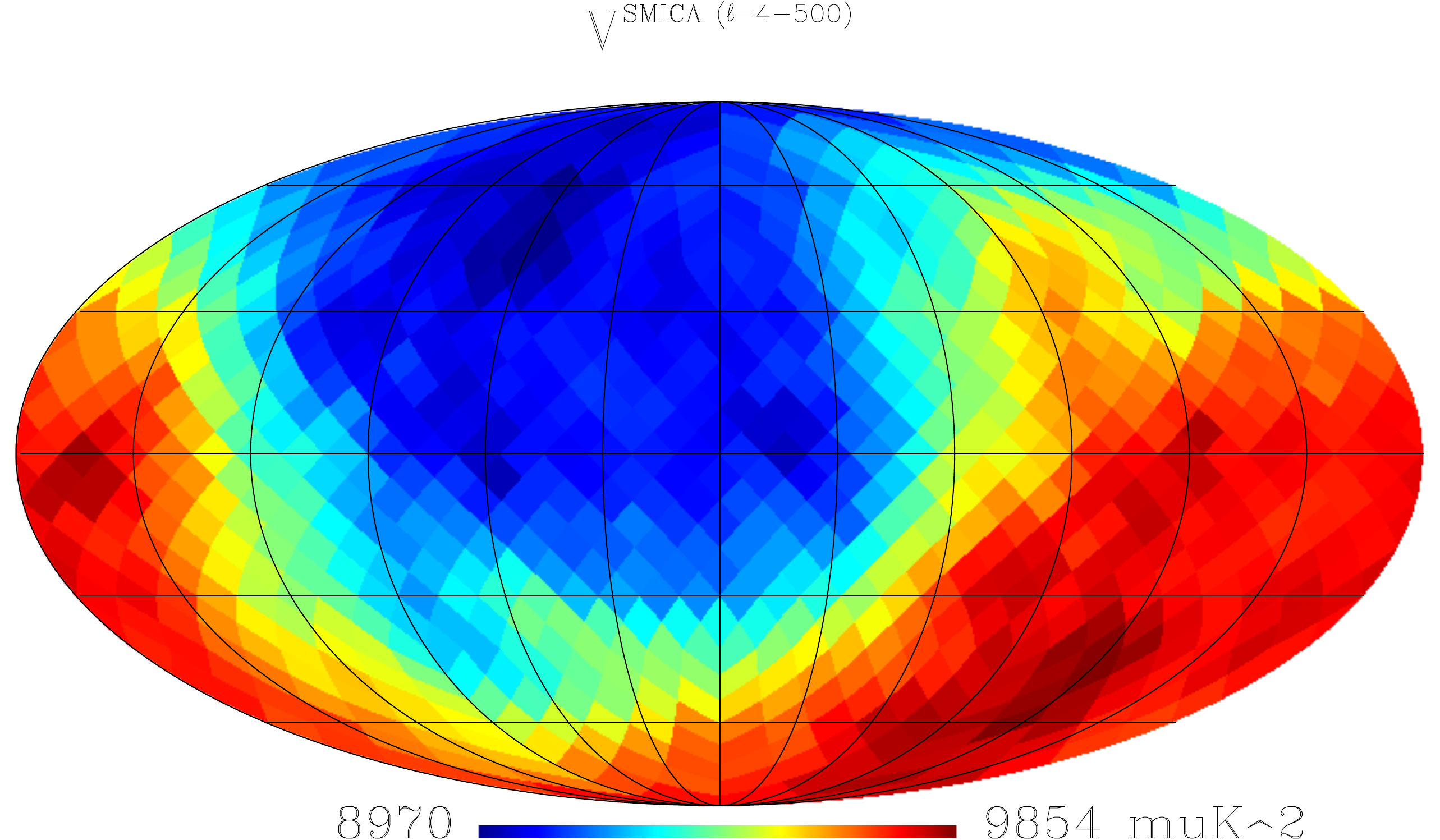}\vspace{1mm}
\includegraphics[scale=0.3,angle=90]{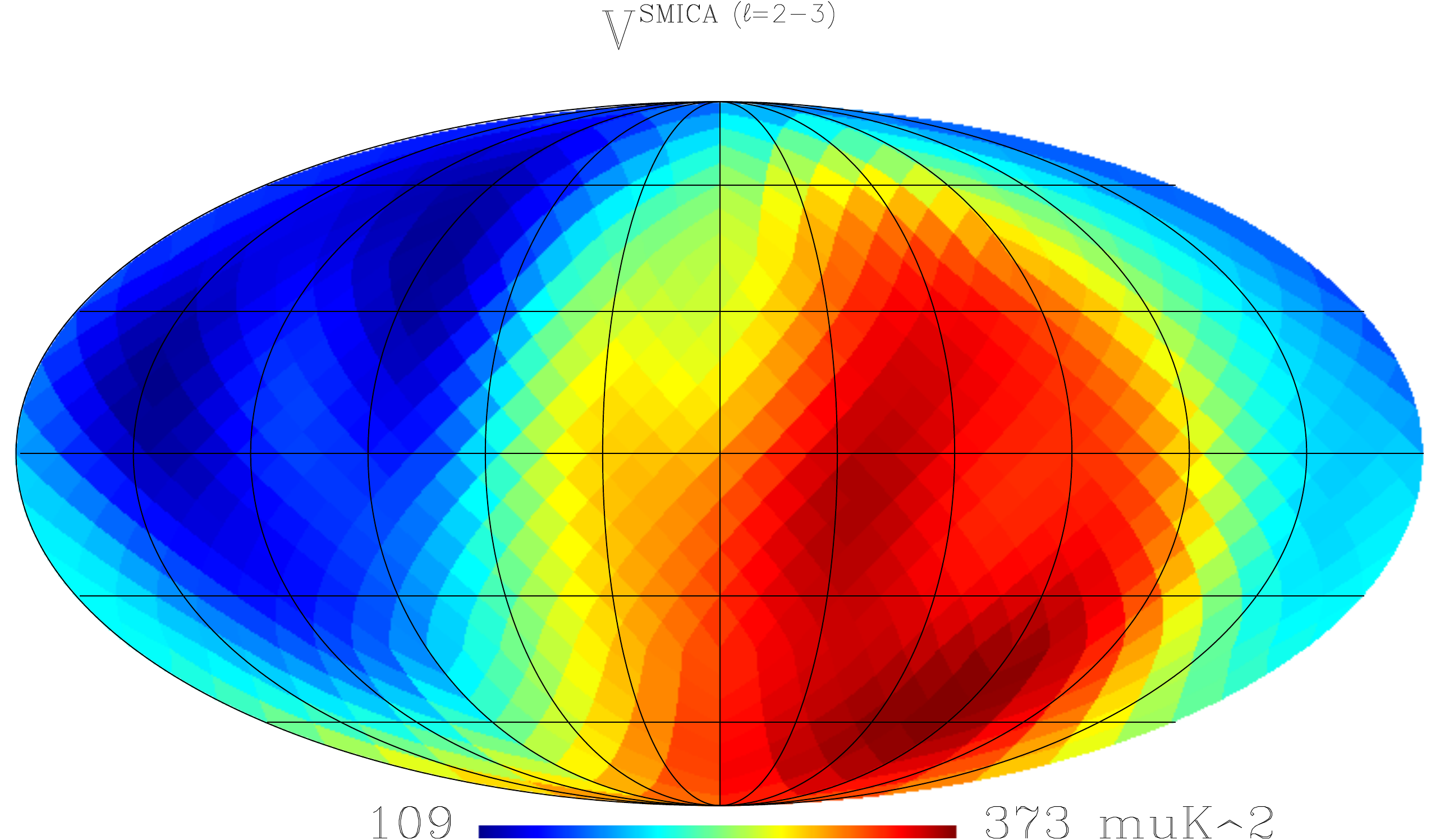}
\caption{\label{fig4} 
Variance dipole direction from the $V^{\mbox{\footnotesize\sc smica}}\!-$map, 
obtained using its {\sc val}mask. In the upper panel we display the case $\ell \in[2, 500]$, in the 
middle panel the case $\ell = [4, 500]$, and in the bottom panel we show the $\ell \in[2, 3]$ case. 
}
\end{center}
\end{figure}

The observed direction of the dipolar variance maps, which appears close to the hemispherical NS-asymmetry~\cite{Land05,Eriksen07}, brings the question about what are the CMB angular scales related to this phenomenon. 
Thus, we apply our variance estimator to the Planck and MC maps containing multipoles 
in a given range: $\ell \in[\ell_\text{min},\ell_\text{max}]$, that is, only contributions in a such angular-scale interval. We consider the {\sc smica} + {\sc val}mask case. The resulting confidence levels and variance dipole directions are summarized in the Table~\ref{table3}. Some of the $V^{\mbox{\footnotesize\sc pla}}\!-$maps corresponding to these angular-scales analyses are shown in Fig.~\ref{fig4}.
We found the following outcomes: 

\begin{itemize}[leftmargin=0.3cm]

\item For angular scales in the interval $\ell\in[2,\, 1,\!000]$, i.e. $\ell_{\text{max}}$=1,000, 
we found that the statistical significance of the variance dipole has only 83.2 \% CL, as compared 
to the dipoles from the $V^{\mbox{\footnotesize\sc g}}\!-$maps, and points in the direction 
$(l, b) \,\simeq\, (245^{\circ},-35^{\circ})$. 
In other words, 168 dipole values, in the set of 1,000 $\{ v_{1}^{\mbox{\footnotesize\sc g}} \}$, 
have a larger value than $v_{1}^{\mbox{\footnotesize\sc smica}}$.

\item For the scales $\ell\in[2,500]$ the variance dipole is, again, not statistically 
significant, 82.8 \% CL; the direction being $(l, b) \,\simeq\, (245^{\circ},-37^{\circ})$.

\item For $\ell \in[2,40]$ we obtain a statistical confidence value of 
71.0 \% CL, with the dipole direction towards $(l, b) \,\simeq\, (250^{\circ},-40^{\circ})$.

\item More interesting, for the CMB scales $\ell \in[41,500]$ the phenomenon is more significant, 
94.3 \% CL, i.e., $\sim 2 \sigma$, with direction towards $(l,b) \,\simeq\, (225^{\circ},-15^{\circ})$, 
close to the NS-asymmetry phenomenon direction.

\item Remarkably, for the angular scales $\ell \in[4,500]$ the variance dipole 
phenomenon is definitely statistically significant:  98.1 \% CL, and the dipolar direction 
$(l,b) \,\simeq\, (220^{\circ},-32^{\circ})$ remains close to the NS-asymmetry phenomenon. 
Notice that removing the $\ell= 2, 3$ multipoles the significance increases, converting it in a 
statistically significant phenomenon: 82.8 \% $\rightarrow$ 98.1\%, contrary to what was 
expected~\cite{Cruz11}.

\item Furthermore, for the case with only the quadrupole ($\ell=2$) plus octopole ($\ell=3$) 
CMB components, we found a low statistical significance: 
$\!$14.0 $\!$\% $\!$CL, with the dipole direction $(l,b) \,\simeq\, (310^{\circ},-25^{\circ})$. 
In addition, the analysis for the largest scales $\ell\in[2,10]$ exhibits again a low significance: 
44.3 \% CL, with dipole direction $(l,b) \,\simeq\, (260^{\circ},-47^{\circ})$. 
\end{itemize}

Contrary to the NS hemispherical asymmetry, where the signal comes from the low CMB 
multipoles (that is, large angular scales; see, e.g.,~\cite{Eriksen07}, and references 
therein), our results show that the power of the variance dipole asymmetry phenomenon 
does not come only from the lowest multipoles ($\ell \le 40$). 
Instead, we observe that the contribution from small scales $\ell \in[41,500]$ is far from being 
negligible. The statistical evidence indicates that the contribution of the CMB multipoles $\ell \in[4,500]$ 
to the variance dipole phenomenon is highly significant: $\sim 2.4 \sigma$.

\begin{table}[h]
\begin{center}
\begin{tabular}{lcc} 
\hline \hline
\ \ CMB angular-scales\ \ & \ \ \ \ \ \ \ \ \  CL (\%) \ \ & \ \ \ \ \ \ $(l, b)$ \ \ \ \\ 
\hline
\ \ \  $\ell \in [2,\,1,\!000]$  & \ \ \ \ \ \ \ 83.2 \ \ \ & \ \ \ \ \ \ \ $(245^{\circ},-35^{\circ})$ \ \ \   \\
\ \ \  $\ell \in [2,500]$    & \ \ \ \ \ \ \  82.8 \ \ \ & \ \ \ \ \ \ \ $(245^{\circ},-37^{\circ})$ \ \ \    \\
\ \ \  $\ell \in [2,40]$      & \ \ \ \ \ \ \  71.0 \ \ \ & \ \ \ \ \ \ \ $(250^{\circ},-40^{\circ})$ \ \ \    \\
\ \ \  $\ell \in [41,500]$  & \ \ \ \ \ \ \  94.3 \ \ \ & \ \ \ \ \ \ \ $(225^{\circ},-15^{\circ})$ \ \ \   \\
\ \ \  $\ell \in [4,40]$      & \ \ \ \ \ \ \ 92.2 \ \ \ & \ \ \ \ \ \ \ $(220^{\circ},-37^{\circ})$ \ \ \     \\
\ \ \  $\ell \in [4,500]$    & \ \ \ \ \ \ \ 98.1 \ \ \ & \ \ \ \ \ \ \ $(220^{\circ},-32^{\circ})$ \ \ \    \\
\ \ \  $\ell \in [2,10]$      & \ \ \ \ \ \ \ 44.3 \ \ \ & \ \ \ \ \ \ \ $(260^{\circ},-47^{\circ})$ \ \ \     \\
\ \ \  $\ell \in [2,3]$        & \ \ \ \ \ \ \ 14.0 \ \ \ & \ \ \ \ \ \ \ $(310^{\circ},-25^{\circ})$ \ \ \     \\
\hline \hline
\end{tabular}
\end{center}
\caption{Statistical angular-scale analyses, showing the confidence level and the variance dipole 
direction for each interval investigated. 
Remarkably, for $\ell \in[4,500]$ the variance dipole phenomenon is 
highly significant: $\sim 2.4 \sigma$. 
Some of the $V^{\mbox{\footnotesize\sc pla}}\!-$maps corresponding to these angular-scales 
analyses are shown in~Fig.\ref{fig4}.} 
\label{table3}
\end{table}

\subsection{Masks and foreground residuals effects on Variance dipole}%
\label{foregrounds}

\begin{figure}
\begin{center}
\hspace{-0.85cm}
\includegraphics[scale=0.48]%
{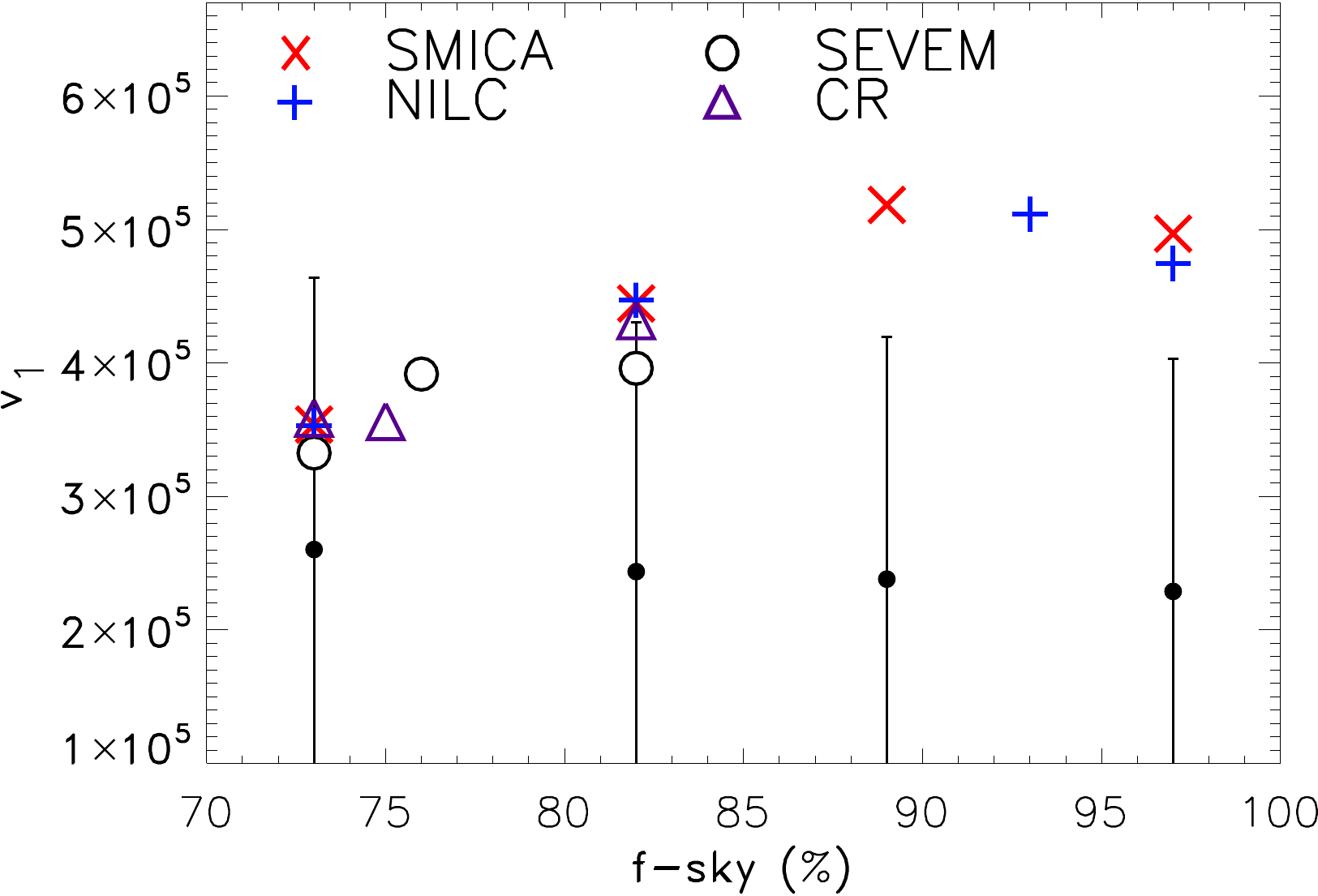}
\caption{\label{fig5} 
Plot of the $v_{1}^{\mbox{\footnotesize\sc pla}}$ and 
$\overline{v}_{1}^{\mbox{\footnotesize\sc g}}$ 
values vs. the $f_{\mbox{\footnotesize sky}}$ (see Table~\ref{table1} for details). 
As observed, given a mask (i.e., $f_{\mbox{\footnotesize sky}}$), all the four 
foreground-cleaned Planck maps have almost equal 
$v_{1}^{\mbox{\footnotesize\sc pla}}$ values. 
\,In other words, all Planck maps produce $V^{\mbox{\footnotesize\sc pla}}\!-$maps 
with the same dipole when the same mask is applied to them.}
\end{center}
\end{figure}

The four foreground-cleaned Planck maps, are expected to be free of 
contamination in the region outside their validation masks (i.e., outside the 
{\sc val}masks)~\cite{PLA-XII}. 
Now we analyse the effect of the cut-sky masks on the variance dipole values obtained from different foreground-cleaned maps. In Fig.~\ref{fig5} it is shown that, for different values of $f_{\mbox{\footnotesize sky}}$, the mean variance dipole from simulations, $\overline{v}_{1}^{\mbox{\footnotesize\sc g}}$, remains almost constant under the different cut-sky masks employed (see Table~\ref{table1}). 
However, this plot also exhibits that the values $v_{1}^{\mbox{\footnotesize\sc pla}}$ decrease when larger masks are used. Two notable informations are concentrated in this interesting plot. The first one points against foreground residuals hypothesis: because all foreground-cleaned maps behaves identically when the same mask 
is used in their analysis, this is an strong indication that foregrounds residuals are absent in these maps. Second, from Fig.~\ref{fig5} there is a clear inference that the intensity of the dipole value from Planck maps is concentrated near the galactic plane, and for this reason much power is cut-off when larger masks are used.

We now investigate the frequency dependence of the $V^{\mbox{\footnotesize\sc pla}}\!-$maps, since it is well-known that galactic foregrounds depend on the electromagnetic 
frequency~\cite{PLA-XII} and therefore their foreground residuals, if present, 
would manifest differently according to the individual frequency of the map in study. 
In accordance with this, and using the most severe cut-sky, i.e., the U73 mask, we produced the $V^{\,\nu}\!-$maps for the individual frequency Planck maps of 70, 100, and 143 GHz. We found that their dipole values are quite similar: 2.44, 2.33, and 
$1.85 \times 10^{5} \mu \mbox{K}^2$, respectively. 
Moreover, the dipole directions are also analogous: 
$(l, b) \,\simeq\, (245^{\circ},-35^{\circ})$, 
$(l, b) \,\simeq\, (240^{\circ},-40^{\circ})$, 
and $(l, b) \,\simeq\, (235^{\circ},-35^{\circ})$, respectively. 
Therefore, the fact that the variance dipoles $v_{1}^{\,\nu}$ for the individual frequency Planck maps, $\nu = 70, 100, 143$ GHz, are very similar, suggests that the galactic foreground residuals are not causing the variance dipole effect. 

Our conclusion is, in accordance with~\cite{PLA-XXIII} but in disagreement with a previous report~\cite{Cruz11}, that galactic foregrounds are unlikely to be the cause of the variance dipole phenomenon. Although this conclusion is not new~\cite{PLA-XXIII}, its confirmation by a different statistical procedure is reassuring. 

To end this subsection, we find interesting to explain why removing the quadrupole and the octopole components from data and simulations increases the statistical significance, instead of decreasing it as claimed in~\cite{Cruz11}. 
The behavior of two quantities are relevant in this examination: the intensity, 
$v_{1}^{\mbox{\footnotesize\sc pla}}$, and the direction, $(l, b)$, of the variance dipole for the angular-scales in analyses. 
Regarding the role of the dipole direction: going from the case $\ell \in[2,500]$ to the case $\ell \in [4,500]$ the dipole direction changes from $(245^{\circ},-37^{\circ})$ to $(220^{\circ},-32^{\circ})$, that is, the net effect on the dipole direction after removing the $\ell = 2 - 3$ components is to point closer the galactic plane region (see Table~\ref{table3}). Regarding the intensity of the dipole: as commented above, from Fig.~\ref{fig5} one deduces that the strength of the dipoles $v_{1}^{\mbox{\footnotesize\sc pla}}$ is concentrated near the galactic plane~\footnote{It was shown in~\cite{Cruz11} that the variance of the CMB data is not stable 
against the Galactic masks used.}, and due to this fact much power is cut-off when larger masks are used. In this way, considering the galactic cuts, like $|b| \simeq 30^{\circ}$, used in~\cite{Cruz11} it is not difficult to understand that the dipole intensity is being cut-off in a larger fraction in the case 
$\ell \in[4,500]$ as compared with the original case $\ell \in[2,500]$. 
Differently from that galactic cuts, the {\sc val}mask used here is not so large near the region 
$(l, b) \sim (220^{\circ},-32^{\circ})$, so in our case the dipole intensity cut is moderate.

It is also worth to illustrate the role of the CMB multipole's intensity in the growth in the statistical 
confidence level when going from $\ell \in[2,500]$ (82.8\% CL) to the case $\ell\in[4,500]$ (98.1\% CL). 
The reason seems to be in the large difference between the low multipole's moments 
(in $\mu K^2$ units) $C_2 \simeq 250$ and $C_3 \simeq 480$ in Planck CMB maps, as 
compared with the corresponding values in $\Lambda$CDM and MCs spectra, namely, 
$C_2^{\Lambda{\text CDM}} = 1158$ and 
$C_3^{\Lambda{\text CDM}} = 545$~\footnote{http://pla.esac.esa.int/pla/aio/planckProducts.html}. 
Because the quadrupole and octopole are notoriously lower for the Planck maps as compared with those of the MC produced from the $\Lambda$CDM spectrum, it makes intuitive sense that the variance dipoles from data and simulations are best compared in the range $\ell \in[4,500]$ than in the interval $\ell \in[2,500]$. 
To test this intuitive argument, we make the following experiment. 
Consider the {\sc smica} map with its $\{ a_{2 m} \}$ and $\{ a_{3 m} \}$ components now multiplied by a numerical factor in such a way that its new quadrupole and octopole moments are $C_2^{\text new} = C_2^{\Lambda{\text CDM}}$ and 
$C_3^{\text new} = C_3^{\Lambda{\text CDM}}$. 
Then, we repeat the variance analysis for this map, considering the angular scales 
$\ell \in[2,500]$, now finding that the statistical confidence level is 98.3 \%, very similar with the 98.1 \% CL found for the case $\ell \in[4,500]$.

\section{Concluding remarks} \label{conclusion}

The Gaussian and statistically isotropic scenario, on which the $\Lambda$CDM concordance model is based, can be rigorously tested with precise CMB data from the four foreground-cleaned Planck maps. Although questions regarding the statistical homogeneity of the universe's large-scale structure wait for future large and deep surveys~\cite{JPAS}, other stimulating questions can be addressed with the 
highly precise CMB data from the Planck satellite.

Using a directional variance estimator, based on the {\em variance} statistical momentum, we performed a statistical analyses of the four foreground-cleaned Planck maps in several angular-scales intervals. In all the intervals investigated our results reveal a net dipolar distribution. In particular, in the angular 
scales $\ell \in[4,40]$ and $\ell \in[41,500]$ the significance is moderate, $\sim 2 \sigma$. Moreover, for the multipoles range $\ell \in[4,500]$, the result is highly significant $\sim 2.4 \sigma$ (see Table~\ref{table3}), with the variance dipole pointing in the direction $(l, b) \,\simeq\, (220^{\circ},-32^{\circ})$, close to the direction of the NS-asymmetry phenomenon.

Additionally, we found that the Planck's variance dipole magnitude gets lower values for larger sky-cut masks, independent of the map analysed. 
This fact is coherent with the result that the variance dipole direction, for all the angular-scale intervals analysed, points relatively near the galactic plane (see Table~\ref{table3}): in such a case, larger masks (i.e., lower $f_{\mbox{\footnotesize sky}}$) cut-off larger regions near this plane where most of the power is located.

Moreover, we found that foreground residuals are absent in our analyses because, considering the same mask, all the foreground-cleaned maps have essentially the same variance dipole value (with a slight dispersion of $\pm 3\%$, see Fig.~\ref{fig3}). 
On the one other hand, this variance dipole stands robust in magnitude and direction against frequency dependence, in the 70, 100, and 143 GHz maps, disfavouring the foreground residuals cause, in agreement with~\cite{PLA-XXIII}. 

Furthermore, an important part of the analyses of the foreground-cleaned Planck maps that validate our results was the robustness tests, where such examinations considered realistic features of the data like the inhomogeneous noise maps and galactic cut-sky masks (information released by the {\em Planck collaboration}~\footnote{Based on observations obtained with Planck (http://www.esa.int/Planck), an ESA science mission with instruments and contributions directly funded by ESA Member States, NASA, and Canada.}). The inhomogeneous noise comes out as a result of the non-uniform way the CMB sky is measured by the Planck probe. In fact, the regions near the ecliptic poles were observed by the probe many more times than others. The pixel's inhomogeneous noise data were released together with the foreground-cleaned Planck 
maps and are crucial to assert its influence in the hemispherical asymmetry 
found in the $V^{\mbox{\footnotesize\sc pla}}\!-$maps. In fact, although this noise has small magnitude, i.e. $|T| \lesssim 18 \, \mu$K at 1$\sigma$ level, the inhomogeneous noise has to be included in the analyses in order to quantify its 
impact in the results, and most importantly, to test the robustness of our outcomes. 
Our results show (see Table~\ref{table2}) that including pixel's inhomogeneous noise in the data analyses does not modifies appreciably our confidence level's calculation listed below. For instance, for the {\sc smica+valmask} map, with and without noise, we obtain: $v_1^{\text smica+noise} / v_1^{\text smica} = 0.985$.

Summarizing, we conclude that our directional variance estimator shows a clear dipolar structure in the four foreground-cleaned and the individual frequency Planck maps, results that appear robust against the component separation algorithms, various Planck masks, map's pixelization parameters, and the addition of inhomogeneous pixel's noise. 
The magnitude of this dipole is highly significant, $\sim 2.4 \sigma$, in the angular scale interval $\ell \in[4,500]$, attaining less significant values in the scales $\ell \in[4,40]$ and $\ell \in[41,500]$ (Table~\ref{table3}). 
We also discover that this significance is not so high in the range $\ell \in[2,500]$ just because the $C_2$ and $C_3$ values in Planck CMB maps are manifestly lower than the corresponding values in the MC maps we use for analyses, which are based on $\Lambda$CDM spectrum. As a matter of fact, if we increase these multipoles values in Planck maps to be equal to those in MC data, we found that the statistical significance in this interval increases from 
82.8\% to 98.3\%, i.e., $ 2.4 \sigma$.

\vspace{4mm}
\begin{acknowledgments}
\noindent
TSP and AB acknowledge the support of Conselho Nacional de Desenvolvimento
Cient\'{\i}fico e Tecnol\'{o}gico (CNPq) -- Brasil. 
We acknowledge the use of the Planck data. 
Some of the results in this paper were derived using the HEALPix 
package~\cite{Gorski05}. 
\end{acknowledgments}

\bibliographystyle{JHEP}
\phantomsection\addcontentsline{toc}{section}{\refname}\bibliography{referencias}

\end{document}